\def\beq{\begin{equation}}
\def\eeq{\end{equation}}
\def\beeq{\begin{eqnarray}}
\def\beeqn{\begin{eqnarray*}}
\def\eeeq{\end{eqnarray}}
\def\eeeqn{\end{eqnarray*}}

                 \def\D{\Delta}

\def\frac#1#2{ {{#1} \over {#2} }}

\newcommand{\unity}{1\kern-.65mm \mbox{\form l}}
\newcommand{\ks}{\mbox{\form l}\kern-.6mm \mbox{\form K}}
\newcommand{\A}{A \raise0.5mm\hbox{\kern-1.8mm /}}
\def\pmb#1{\leavevmode\setbox0=\hbox{$#1$}\kern-.025em\copy0\kern-\wd0
\kern-.05em\copy0\kern-\wd0\kern-.025em\raise.0433em\box0}
\def\D{\hbox{\hbox{${D}$}}\kern-1.9mm{\hbox{${/}$}}}
\def\kbar{\hbox{$k$}\kern-0.2true cm\hbox{$/$}}
\def\nbar{\hbox{$n$}\kern-0.23true cm\hbox{$/$}}
\def\pbar{\hbox{$p$}\kern-0.18true cm\hbox{$/$}}
\def\nhbar{\hbox{$\hat n$}\kern-0.23true cm\hbox{$/$}}

\documentstyle[epsfig,preprint,aps,floats,amssymb]{revtex}
\begin{document}
\draft
\newfont{\form}{cmss10}
\title{Cutting rules and perturbative unitarity of noncommutative electric-type field theories from string theory}  
\author{Alessandro Torrielli$^1$}
\address{$^1$Dipartimento di Fisica ``G.Galilei", Via Marzolo 8, 35131
Padova, Italy\\
INFN, Sezione di Padova, Italy \\
\texttt{(torrielli@pd.infn.it)}}
\maketitle
\begin{abstract} 
We discuss the breakdown of perturbative unitarity of noncommutative quantum field theories in electric-type background in the light of string theory. We consider the analytic structure of string loop two-point functions using a suitable off-shell continuation, and then study the zero slope limit of Seiberg and Witten. In this way we pick up how the unphysical tachyonic branch cut appears in the noncommutative field theory. We briefly discuss discontinuities and cutting rules for the full string theory amplitude and relate them to the noncommutative field theoretical results, and also discuss the insight one gains into the magnetic case too.\end{abstract}

DFPD 02/TH 18.

PACS numbers: 11.25.Db, 11.10.Kk, 11.55.Bq
\section{Introduction}
The problem of the perturbative unitarity of noncommutative quantum field theories has been faced since the appearance of these theories as an effective description for open string theory amplitudes in an antisymmetric constant background \cite{sw}. The absence of a straightforward  hamiltonian formulation for the case of an electric-type noncommutativity (in which the time variable is involved) and the bizarre dynamical features of its scattering amplitudes \cite{sst} have soon casted doubts on the consistency of this kind of theories. One of the first confirmations of the breakdown of perturbative unitarity was given in \cite{gm} for a noncommutative scalar theory: Cutkoski's rules were found to hold only in the magnetic case; in the electric case an additional tachyonic branch cut is present \footnote{The case of a light-like noncommutativity is peculiar and was discussed in \cite{agm}}. An analogous result was found in \cite{bgnv} for noncommutative gauge theories: here again the cutting rules hold if the noncommutativity does not involve the time variable, otherwise new intermediate tachyonic states call for being inserted in order to explain the analytic structure of the vacuum polarization tensor \footnote{For a discussion on the role of UV/IR mixing in relation to unitarity in NC field theories see \cite{clz}}. These new possible states were considered in \cite{abz}: the claim is that unitarity can be recovered provided one add the appropriate intermediate exchanges. This is related to the fact that in the electric case the Seiberg-Witten limit does not succeed in decoupling the open massive string states, as well as the closed string sector. The best one can do in general in this situation is to send the electric field at its critical value, ending with a theory of only open strings (NCOS), namely decoupling all closed strings but keeping the open massive tower of states (\cite{ssto,gmms}, see also \cite{kp,k}). 

From the point of view of the behaviour of open strings in electric backgrounds the situation is well understood \cite{ft,acny,clny,b,n,bp,ba}: there is a critical value of the field beyond which the string becomes unstable. In particular an electric field along the string can strecht it and balance its energy, reducing its effective tension to zero, making pair production possible from the vacuum without energy loss. The limit of Seiberg and Witten precisely forces the electric field to overcome this critical value; therefore the corresponding field theory describes an unstable string. There is no way to reach a complete decoupling without overcoming this critical point, therefore the resulting low energy truncated effective theory is necessarily tachyonic, even if the full string theory from which one starts is taken in the region of stability.
      
Our aim is to make this situation clear from the point of view of the analytical structure of the full string theory amplitude, namely we want to reproduce the result of \cite{gm} from the string two point function, precisely looking at what happens to the branch cuts when one performs the Seiberg-Witten limit. We want to explain why it is necessarily problematic to go down from the string theory amplitudes onto the field theoretical ones from the point of view of the branch cuts they both exhibit. We see that the string theory two-point amplitude, when viewed as a function of the squared momentum $p^2$ of the external leg, has two positive branch cuts in the complex plane below the critical value of the electric field. Our off-shell continuation of the string amplitude is performed in the spirit of Di Vecchia et al.\cite{div1,div2,div3} (see also \cite{bern1,bern2,bern3}) and of \cite{bcr}, which is the way we use to refer the result of \cite{gm} to the string theory calculation. One of the two branch cuts is parameterized by a quantity that changes sign when the electric field overcomes its critical value, and this situation is reflected in the analytical structure of the field theory amplitudes. This enforce the convinction that the anomalous spectral features of NC quantum field theory have a correspondence at the level of the related string model, and we consider it remarkable that the breaking of unitarity of the effective field theory corresponds to the breaking of stability of the original theory. Otherwise stated, the breaking of perturbative unitarity in the NC electric-type field theory is due to the fact that its amplitudes are precisely equal to the amplitudes of an open string in an electric background, when one forces them in the limit of Seiberg-Witten: this is an illicit operation, since an effective field theory for this situation does not exist, and the result is not a field theory, which is what happens to time-like noncommutative QFT.

We further comment on the relevance of our kind of analysis in the magnetic case too, where the existence of the Seiberg-Witten decoupling limit and the formal recovering\cite{gm} of the cutting rules in the NC limit does not reveal the nature of the infrared singularity, extraneous to the elementary massive scalar particles one starts with. We relate it to an off-shell manifestation of the closed string sector, related to the above mentioned problem. We end with a calculation of the discontinuities across the branch cuts of the full string amplitude in opportune approximations, and with some remarks.  
\section{Analysis of the branch cuts}
Let us begin considering one-loop tachyon amplitudes in bosonic open string theory in the presence of a constant $B$-field living on a D$p$-brane \footnote{For the derivation of amplitudes for photon-vertex operators the reader is referred to \cite{bcr,gkmrs}}. These amplitudes have already been studied \cite{ad,kl,bcr,l,crs}, in particular in \cite{ad} it was shown that the noncommutative effective field theory on the brane in the zero-slope limit was precisely NC ${\phi}^3$. This was also one of the cases examined in \cite{gm} where the breakdown of cutting rules was found. 

We therefore start from writing the string amplitude for a generic dimension $p$ of the brane, governed by the sigma model action 
\begin{eqnarray}
\label{1}
S={{1}\over{4\pi {\alpha}'}}\int_{C_2}d^2 z \, (g_{ij} {\partial}_a X^i {\partial}^a X^j - 2 i \pi {\alpha}' B_{ij} {\epsilon}^{ab}{\partial}_a X^i {\partial}_b X^j).     
\end{eqnarray}
The string world-sheet is conformally mapped into the cilinder $C_2 = \{0\leq \Re w\leq 1, w = w + 2 i \tau$\}; $\tau$ is the modulus of the cilinder, the metric $g$ and the antisymmetric tensor $B$ are constant closed-string sector backgrounds. The indices $i$ and $j$ live on the brane: since outside the brane a constant $B$ can be gauged away, we are interested only in the brane action. The propagator we will adopt in this situation, with the new boundary conditions imposed by the $B$-term in (\ref{1}), has been constructed in \cite{acny},\cite{clny},\cite{ad}. If one sets $w=x+iy$, the relevant propagator on the boundary of the cilinder ($x=0,1$) can be written as \cite{ad}
\begin{eqnarray}
\label{2}
G(y,y')={{1}\over{2}}{\alpha}' g^{-1} \log q - 2 {\alpha}' G^{-1} \log \Big[q^{{1}\over {4}} \, {\vartheta_4}({{|y-y'|}\over {2 \tau}}, {{i}\over {\tau}}) / D(\tau)\Big],\, \, \, \, \, \, \, \, \,     x\neq x', 
\end{eqnarray}
\begin{eqnarray}
\label{3}
G(y,y')=\pm {{1}\over{2}} i \theta \, {\epsilon}_{\perp} (y-y') - 2 {\alpha}' G^{-1} \log \Big[{\vartheta_1}({{|y-y'|}\over {2 \tau}}, {{i}\over {\tau}}) / D(\tau)\Big],\, \, \, \, \, \, \, \, \,     x=x',
\end{eqnarray}  
where $q=e^{-{{\pi}\over{\tau}}}$, $\pm$ correspond to $x=1$ and $x=0$ respectively, and ${\epsilon}_{\perp} (y) = sign(y) - {{y}\over{\tau}}$.
The open string parameters are as in \cite{sw}:
\begin{eqnarray}
\label{4}
G=(g-2\pi {\alpha}' B)g^{-1} (g+2\pi {\alpha}' B)
\end{eqnarray}
is the open string metric, and
\begin{eqnarray}
\label{5}
\theta=-{(2\pi{\alpha}' )}^2 {(g+2\pi {\alpha}' B)}^{-1}B {(g-2\pi {\alpha}' B)}^{-1}
\end{eqnarray}
is the noncommutativity parameter. ${\vartheta_{4,1}}(\nu, \tau )$ are Jacobi theta functions, while $D(\tau)={\tau}^{-1} {[\eta({i\over \tau})]}^3$, where $\eta$ is the Dedekind eta function\cite{p}.

With this propagator and the suitable modular measure, the amplitude for the insertion of $N$ tachyonic vertex operators at $x=1$ and $M-N$ at $x=0$ has been calculated \cite{ad} 
\begin{eqnarray}
\label{6}
A_{N.M}&=&{\cal{N}}{(2\pi)}^{d} {({\alpha}')}^{\Delta} {G_s}^M \int_0^{\infty} {{d\tau}\over{\tau}} {\tau}^{-{{d}\over{2}}} {[\eta(i\tau)]}^{2-d} q^{{{1}\over{2}}{\alpha}' Kg^{-1} K}  \nonumber \\
&&\times \Big(\prod_{a=1}^M \int_0^{y_{a-1}} dy_a \Big) \prod_{i=1}^N \prod_{j=N+1}^M {\Bigg[ q^{{1}\over {4}} \, {\vartheta_4}({{|y_i - y_j|}\over {2 \tau}}, {{i}\over {\tau}}) / D(\tau)\Bigg]}^{2 {\alpha}' k_i G^{-1} k_j } \nonumber \\
&&\times \prod_{i<j=1}^N e^{- {{1}\over{2}}i {\epsilon}_{\perp} (y_i - y_j)k_i \theta k_j} {\Bigg[ {\vartheta_1}({{|y_i -y_j|}\over {2 \tau}}, {{i}\over {\tau}}) / D(\tau)\ \Bigg]}^{2 {\alpha}' k_i G^{-1} k_j }\nonumber \\
&&\times \prod_{i<j=N+1}^M e^{{{1}\over{2}}i {\epsilon}_{\perp} (y_i - y_j)k_i \theta k_j} {\Bigg[ {\vartheta_1}({{|y_i -y_j|}\over {2 \tau}}, {{i}\over {\tau}}) / D(\tau)\ \Bigg]}^{2 {\alpha}' k_i G^{-1} k_j } + noncycl. \, perm.
\end{eqnarray}
Here ${\cal{N}}$ is the normalization constant, $d=p+1$, $\Delta = M {{d-2}\over{4}} - {{d}\over {2}}$, $G_s$ is the open string coupling constant, $K=\sum_{i=1}^N k_i$ is the sum of all momenta associated with the vertex operators on the $x=1$ boundary, and ${y_0} = 2\tau$. One can see that, when $N,M\neq0$, this amplitude corresponds to nonplanar graphs: this is most easily realized when mapping the cilinder into an annulus whose internal boundary corresponds, say, at $x=0$ and the external one at $x=1$. We have also omitted in the amplitude the global delta function due to momentum conservation and the traces of the Chan-Paton matrices.

We want to discuss the case of $N=1$, $M=2$, that is a diagram with two vertex inserted on opposite boundaries, which, in the field theory limit, will become the nonplanar contribution to the two-point function. This was the first example in \cite{ad}, section 2.2: we will give here a more detailed derivation, using a suitable rescaling and in order to set the asymptotic formulae useful for the subsequent analysis. Nonplanar diagrams are the ones in which the dependence on noncommutativity does not factor out the loop integral and therefore they are substantially different from their commutative counterpart\cite{mvs}; they are indeed the diagrams we are interested in. We rescale $t=2\pi {\alpha}' \tau$ and $\nu_{1,2} ={{y_{1,2}}/ {2 \tau }}$. After this we set $\nu_{2} =0$ to fix the residual invariance. The result we obtain is 
\begin{eqnarray}
\label{7}
A_{1.2}&=&{\cal{N}} {G_s}^2 \, 2^{{3 d} \over 2}{\pi}^{{{3 d}\over{2}}-2} {\alpha '}^{{{d}\over{2}}-3} \int_0^{\infty} dt \, t^{1-{{d}\over{2}}} \, {\Big[ \eta({{i t}\over{2\pi {\alpha '}}})\Big] }^{2-d}\times \nonumber \\
&&\, e^{- {{{\pi}^2}{\alpha '}^2 \over {t}} kg^{-1} k} 
\int_0^1 d\nu {\Bigg[{{e^{-{{{\pi}^2}{\alpha '}\over{2t}}} {\vartheta_4}(\nu, {{2 \pi i {\alpha '}}\over {t}})} \over { {{2 \pi{\alpha '}}\over{t}} {[\eta({2\pi i {\alpha '}\over {t}})]}^3 }}\Bigg]}^{- 2 {\alpha '} kG^{-1} k},
\end{eqnarray}
$k$ being the external momentum. 

Now we perform the zero-slope limit ${\alpha}' \to 0$, following the lines sketched in the introduction, keeping $t$, $\nu$, $\theta$ and $G$ fixed: this can be done setting ${\alpha}' \sim {\epsilon}^{1\over 2}$ and the closed string metric $g\sim \epsilon$, and then sending $\epsilon \to 0$ \cite{sw}. The formulae for the asymptotic values of the functions in the integrand are as follows: from the expression $\eta (s) = x^{1\over 24}\prod_{m=1}^{\infty} (1 - x^m)$ where $x=\exp [2 \pi i s]$, we deduce the behaviour $$\eta ({{i t}\over{2\pi {\alpha '}}}) \sim \exp [-{t\over{24 \alpha '}}];$$ by using the property that $\eta (s) = {(- i s)}^{- {{1}\over {2}}} \eta (- {{1}\over {s}})$ we get $$\eta({2\pi i {\alpha '}\over {t}})\sim {({2\pi {\alpha '}\over {t}})}^{-{{1}\over {2}}} \exp [-{t\over{24 \alpha '}}],$$ and finally, since one has the property ${\vartheta_4}(\nu, \tau ) = {(- i \tau)}^{- {{1}\over {2}}} \exp [- {{i \pi {\nu }^2 }\over {\tau}}] \, {\vartheta_2}({{\nu }\over {\tau}},- {{1}\over {\tau}})\, $, and $\, {\vartheta_2}(\rho, \sigma)=\sum_{n=-\infty}^{\infty} e^{\pi i \sigma {(n + {1\over 2})}^2 + 2 \pi i \rho (n + {1\over 2})}$, we obtain $${\theta_4}(\nu, {{2 \pi i {\alpha '}}\over {t}})\sim {({2\pi {\alpha '}\over {t}})}^{-{{1}\over {2}}} e^{{t\over {2 \pi {\alpha '}}}[- {{\pi}\over {4}} + \pi (\nu - {\nu}^2)]}.$$

The field theory coupling constant is related to the open string parameters as $g_f \sim G_s \, {{\alpha}'}^{{{d-6}\over {4}}}$ and the mass is ${m^2} = {{2-d}\over {24 {\alpha}'}}$. Furthermore in the Seiberg-Witten limit $g^{-1}=- {{1}\over {(2 \pi {\alpha}')}^2} \theta G \theta$. Putting everything together, one obtains \cite{ad}
\begin{eqnarray}
\label{8}
A_{1.2}\sim{g_f}^2 \int_0^{\infty} dt \, t^{1-{{d}\over{2}}} \, e^{- m^2 t \, + \, k\theta G\theta k/4 t}\int_0^1 d\nu \, e^{- t\, \nu (1 - \nu )\, k G^{-1} k},
\end{eqnarray}
where some unimportant constants have been omitted. This is exactly the expression for the two-point function in the noncommutative $\phi^3$ theory, and we see that the dimension of the brane worldvolume may serve as a parameter of dimensional regularization for the effective field theory. An important point is that now we choose the brane to be actually a string, therefore $d=2$. This case is peculiar since the tachyon mass squared goes to zero from below (if $d>2$ the effective theory is naturally tachyonic, ${m^2} = {{2-d}\over {24 {\alpha}'}}$, we recall that this represents a two-tachyon amplitude in string theory).  

In this situation the limit (\ref{8}) equals the nonplanar amplitude of a massless scalar in two-dimensional NC-$\phi^3$ theory
\begin{eqnarray}
\label{10}
A_{1.2}(d=2)\sim{g_f}^2 \int_0^{\infty} dt \int_0^1 d\nu \, e^{- t \, \nu \, (1 - \nu )\, kG^{-1} k\,  + \, {1\over {4t}}k\theta G\theta k}.
\end{eqnarray}
In two dimensions $\theta$ is proportional to the antisymmetric tensor: ${\theta}^{\mu \nu }=\theta {\epsilon}^{\mu \nu }$. We are therefore in the right situation to study the effects of electric-type backgrounds \footnote{We also notice that the noncommutative nonplanar diagram equals the planar one if we set $\theta=0$}. The open string metric is proportional to the flat Minkowski metric ${\eta}_{\mu \nu }$, and we define a constant G such that $G^{-1} k^2 = k_{\mu} {[G^{-1}]}^{\mu \nu} k_{\nu}$, where $k^2$ is the usual Minkowski invariant. The amplitude (\ref{10}) can then be written as 
\begin{eqnarray}
\label{11}
A_{1.2}(d=2)\sim{g_f}^2 \int_0^{\infty} dt \int_0^1 d\nu \, e^{- k^2 [G^{-1} t \, \nu (1 - \nu ) - {{ \theta}^2 \over {4 G^{-1} t}}]}.
\end{eqnarray}
Already in the commutative case such massless scalar theories have severe infrared problems. Therefore we add a small positive mass as a regulator and get 
\begin{eqnarray}
\label{12}
A_{1.2}(d=2)\sim{g_f}^2 \int_0^{\infty} dt \int_0^1 d\nu \, e^{- {m_0}^2 t - k^2 [G^{-1} t \, \nu \, (1 - \nu ) - {{ \theta}^2 \over {4 G^{-1} t}}]}.
\end{eqnarray}
Setting for the moment $G=1$ for simplicity, we see that the integral in (\ref{12}) is convergent in the strip $\{- 4 {m_0}^2 <\Re k^2<0\}$   
\begin{eqnarray}
\label{13}
A_{1.2}(d=2)\sim \int_0^1 d\nu \Bigg[ - {\theta}^2 k^2 {{K_1\Big( \sqrt{- {\theta}^2 k^2 ({m_0}^2 + k^2 \nu (1 - \nu))}\Big)}\over {\sqrt{- {\theta}^2 k^2 ({m_0}^2 + k^2 \nu (1 - \nu))}}}\Bigg],
\end{eqnarray}
and after analytic continuation it defines an analytic functions with two branch cuts $\{\Re k^2 <-4 {m_0}^2\}$ and $\{\Re k^2>0\}$. One of these is necessarily tachyonic and cutting rules are invalidated. If we send the regulator to zero, the two branch points get closer and closer and eventually coalesce.

Let us analize now the full string theory diagram. We refer to the remarks made in the introduction about the spirit we adopt in continuing the string amplitude off-shell (see Di Vecchia et al.\cite{div1,div2,div3}, see \cite{bern1,bern2,bern3}, and \cite{bcr} in the noncommutative context), taking it as the natural tool to interpret the features of the field theory limit, as shown by the concrete mathematical correspondence. The complete string amplitude, before performing any limit, in the case of $d=2$, is easily found from (\ref{7}) to be
\begin{eqnarray}
\label{9}
A_{1.2}(d=2)={\cal{N}} {G_s}^2 {{8\pi}\over {{{\alpha}'}^2}}\int_0^{\infty} dt \int_0^1 d\nu \, e^{- {{{\pi}^2}{\alpha '}^2 \over {t}} kg^{-1} k} {\Bigg[{{e^{-{{{\pi}^2}{\alpha '}\over{2t}}} {\vartheta_4}(\nu, {{2 \pi i {\alpha '}}\over {t}})} \over { {{2 \pi{\alpha '}}\over{t}} {[\eta({2\pi i {\alpha '}\over {t}})]}^3 }}\Bigg]}^{- 2 {\alpha '} kG^{-1} k}
\end{eqnarray}
We set consistently $g_{\mu \nu } = g {\eta }_{\mu \nu }$ and $B_{\mu \nu } = B {\epsilon }_{\mu \nu }$. A simple calculation shows that $G_{\mu \nu }={{1}\over {g}}[g^2 - {(2 \pi {{\alpha '} B)}^2}] {\eta }_{\mu \nu }$. We can therefore rewrite the string amplitude as   
\begin{eqnarray}
\label{14}
A_{1.2}(d=2)&=&{\cal{N}} {G_s}^2 {{8\pi}\over {{{\alpha}'}^2}}\int_0^{\infty} dt \int_0^1 d\nu \, \exp [-k^2 {{{\pi}^2 {{\alpha '}^2 }}\over {g \, t}}] \nonumber \\
&&\times \exp \Bigg[ {{- 2{\alpha '} g k^2 }\over {[g^2 - {(2 \pi {\alpha '} B)}^2]}}\log [{{e^{-{{{\pi}^2}{\alpha '}\over{2t}}} {\vartheta_4}(\nu, {{2 \pi i {\alpha '}}\over {t}})} \over { {{2 \pi{\alpha '}}\over{t}} {[\eta({2\pi i {\alpha '}\over {t}})]}^3 }}] \Bigg].  
\end{eqnarray}
There are two regions we have to study in order to evaluate the convergence of the integral over $t$ in (\ref{14}), namely the ones close to the two extremes. We use again the asymptotic expressions we have reported above for the elliptic functions, and the duality for the function $\eta$ which relates the open and the closed channels. At small $t$, the integrand behaves as 
\begin{eqnarray}
\label{15}
{\Bigg( {{t}\over {2 \pi \alpha '}}\Bigg) }^{- 2 \alpha ' k^2 {{g}\over {[g^2 - {(2 \pi {\alpha '} B)}^2}]}} \exp \Big( - k^2 {{{\pi}^2 {{\alpha '}^2 }}\over {g \, t}}\Big);
\end{eqnarray}
therefore we have convergence for $g \Re k^2 >0$ and a branch cut along the opposite axis. At large $t$ the argument of the $\log$ behaves like $e^{{t \nu (1 - \nu )}\over {2 {\alpha '}}}$; therefore the $\log$ actually produces a linear term, with a net result
\begin{eqnarray}
\label{16}
\exp \Big(  - k^2 t \, {{g}\over {[g^2 - {(2 \pi {{\alpha '} B)}^2}]}}\, \nu (1 - \nu ) \Big).
\end{eqnarray}
We see that the branch cut is parameterized by the quantity ${{g}\over {[g^2 - {(2 \pi {{\alpha '} B)}^2}]}} = {{1}\over {g}} {{1}\over {1 - {\tilde E}^2}}$ where we have defined the effective electric field ${\tilde E} = {2 \pi {{\alpha '} B }\over g}$. This is exactly the ratio $E\over E_{cr}$ of \cite{ssto} which discriminates the stability of the string in this background: $1 - {\tilde E}^2$ must be positive in order to avoid tachyonic instability. This condition is precisely what we find from cutting rules: if this parameter is positive, the region of convergency is $g \Re k^2 >0$, and the branch cut is superimposed to the one coming from the small $t$ analysis \footnote{Because of the small $\nu$ infrared singularity, $d=2$ is quite peculiar, see next section for specifications on this point}; therefore the total amplitude exhibits a single ``physical'' branch cut. If, on the contrary, this parameter becomes negative, as it does in the Seiberg-Witten limit because $g^2 \sim {\epsilon}^2$ and ${\alpha '}^2 \sim\epsilon$, a new branch cut appears on the opposite axis, which is just the unphysical one we find in the limit.    
\section{Calculation of discontinuities and cutting rules}
In this section we perform some explicit calculations of the discontinuities across the branch cuts we have found in the previous analysis, and compare them with the related tree level phase space according to the cutting rules, in order to gain more insight into the correspondence between the string theory amplitude and the noncommutative field theory one. When the electric field overcomes its critical value we have seen that already at the level of the full open string amplitude a tachyonic branch cut appears: no phase space of particles belonging to the actual string spectrum will be able to reproduce the discontinuity across it, therefore the cutting rules are manifestly violated. Therefore we will restrict our subsequent analysis to the case of $|\tilde{E}| < 1$. We recall that the open string spectrum on the $Dp$-brane under consideration is determined by $m^2 = {{N - a}\over {\alpha '}}$, where $a = {{d-2}\over {24}}$ and $d = p + 1$ is the space-time dimension of the brane over which the indices $i,j$ in Eq.(\ref{1}) run. 

When the electric field is smaller than the critical value, we found a single physical branch cut for the string. The amplitude in the Seiberg-Witten limit manifests however an unphysical branch cut because this limit forces $|\tilde{E}|$ to become larger than $1$. Again in the limit there is no hope to have an optical theorem, which reproduces the conclusion of \cite{gm}. This would not be the case, could we start with a magnetic-like theory (which is impossible in $d=2$). The noncommutative field theory analysis of cutting rules was performed in \cite{gm}. Focusing on the full string theory result, an analysis of factorization of the loop amplitude for a general $M$-point function in the presence of the $B$-field was carried out in \cite{ad} \footnote{The reader is also referred to the fundamental treatment of \cite{gsw}}. Their analysis showed the appearance of the closed string channel as usual when ${{t}\over {2 \pi \alpha '}} = \tau \to 0$. This was obtained in the standard way of putting all external momenta on shell, and using momentum conservation, such that $\sum_{i<j=1}^{M} 2 \alpha ' k_i G^{-1} k_j = - M$. The analysis of singularity is then performed in the Mandelstam variables. The two point function is quite peculiar in this respect, as we have already remarked above, since it depends on the only invariant $k^2$. As in the previous section, we will perform an off-shell analysis on $k^2$, and will start directly from the formula for the two point function. 

The amplitude (\ref{9}) is quite complicated, moreover it exhibits an infrared singularity in the large-$t$ corner of integration when $\nu (1 - \nu )$ is very small, as one can see looking at (\ref{16}): integration over $t$ leads to a logarithmically divergent integral. This is peculiar to the two dimensions. In fact we choose to consider the variable $d$ as a complex variable, and go to a strip in the $d$ complex plane in which we do not have this problem, treating the dimension of the worldvolume of the brane as a parameter of dimensional regularization. The amplitude we study now is therefore Eq.(\ref{7}), and the same procedure we used to single out the asymptotic behaviours (\ref{15}) and (\ref{16}) leads us to the following results: at small $t$, 
\begin{eqnarray}
\label{17}
{\Bigg( {{t}\over {2 \pi \alpha '}}\Bigg) }^{2 \alpha ' k^2 {{|g|}\over {[g^2 - {(2 \pi {\alpha '} B)}^2}]}} \exp \Big( k^2 {{{\pi}^2 {{\alpha '}^2 }}\over {|g| \, t}} - {{{\pi}^2 {{\alpha '}} (2 - d)}\over {6 t}}\Big),
\end{eqnarray}
where we have assumed $g$ negative. We select the strip $(2 - d)>0$, which corresponds to $m^2={{(2-d)}\over {24\alpha '}}>0$. The branch cut is for $\Re k^2 > 4 m^2 |g|$. At large $t$ we have the behaviour 
\begin{eqnarray}
\label{18}     
t^{1-{{d}\over {2}}} \, \, \exp \Big( k^2 t \, {{|g|}\over {[g^2 - {(2 \pi {{\alpha '} B)}^2}]}}\, \nu (1 - \nu )  - m^2 t \Big).
\end{eqnarray}
The branch cut is for $\Re k^2 > 4 m^2 [g^2 - {(2 \pi {{\alpha '} B)}^2}] / |g|$ where we recall that we restrict the analysis to $g^2 - {(2 \pi {\alpha '} B)}^2 > 0$. As long as $B\neq 0$ this branch cut starts below the previous one, being $[g^2 - {(2 \pi {{\alpha '} B)}^2}] / {|g|^2} < 1$. Now we see that, integrating $t$ in this region, and then integrating $\nu$, does not give any problem. We remark that, going to complex dimensions $2+(d-2)$, we have chosen to keep the antisymmetric field still of electric type, and, together with the momentum, in the $2$ dimensions with zero components in the other $d-2$ ones. The main change is therefore the appearance of a small positive mass which regulates the integral over $\nu$. 

Our results show, therefore, that the two branch cuts of the string amplitude, which are superimposed in $d=2$, are separated in the strip $d<2$. They have also different nature:  the first one, coming from the corner of small $t$ corresponding to the internal radius of the annulus becoming negligible, is driven by the closed string metric $g$, while the second one, coming from the corner in which the modulus $t$ of the annulus is very large, is driven by the open string metric $G$, and starts well below the first one. 

Evaluating the discontinuities from the full analytic formulae is quite a formidable task, therefore we choose to perform the calculation in the approximation in which the momentum of the incoming external tachyon is just above the starting point of the lower cut of a very small amount. In this situation the higher cut is not yet active, as well as the higher string levels, which start from $m^2 = {{1 - {{d-2}\over {24}}}\over {\alpha '}}$: we break the integration region in order to separate small $t$ from large $t$, then only this second piece develops a discontinuity, the first one resulting in an entire function of the momentum. Here $\alpha '$ is finite, but the momentum is at the threshold, therefore only very high $t$'s are important, and the discontinuity is the one of the field theory-like expression 
\begin{eqnarray}
\label{19}
{\cal{N}} {G_s}^2 \, 2^{{3 d} \over 2}{\pi}^{{{3 d}\over{2}}-2} {\alpha '}^{{{d}\over{2}}-3} \int_0^1 d\nu \, \int^{\infty} dt \, t^{1-{{d}\over{2}}} \, \exp \Big( k^2 t \, {{|g|}\over {[g^2 - {(2 \pi {{\alpha '} B)}^2}]}}\, \nu (1 - \nu )  - m^2 t \Big).  
\end{eqnarray}
 Integrating over $t$ we exploit the freedom we have in choosing the lower extreme, which we fix at ${{1}\over {\gamma}}$, $\gamma = |g| / [g^2 - {(2 \pi {{\alpha '} B)}^2}]$, and get
\begin{eqnarray}
\label{20}
{\cal{N}} {G_s}^2 \, 2^{{3 d} \over 2}{\pi}^{{{3 d}\over{2}}-2} {\alpha '}^{{{d}\over{2}}-3} {\gamma}^{{{d}\over {2}} - 2} \int_0^1 d\nu \, {{\Gamma [\omega, {\mu}^2 - k^2 \nu (1 - \nu )]}\over {{[{\mu}^2 - k^2 \nu (1 - \nu )]}^{\omega }}}, 
\end{eqnarray}
where $\Gamma$ is the incomplete gamma function, $\omega = 2 - (d/2)$, and $\mu^2 = m^2 / \gamma$. The discontinuity is found to be of a logarithmic type, and, using transformation properties of $\Gamma$,  it is explicitly evaluated near the threshold
\begin{eqnarray}
\label{21}
&&Disc A_{1.2} \propto \nonumber \\
&i& {\cal{N}} {G_s}^2 \, 2^{{3 d} \over 2}{\pi}^{{{3 d}\over{2}}-1} {\alpha '}^{{{d}\over{2}}-3} {\gamma}^{{{d}\over {2}} - 2} {{\Gamma [1/2]}\over {\Gamma [(3/2) - \omega ]}} {\left( {{k^2} \over {4}}\right) }^{- \omega } {\left( 1 - {{4 \mu^2}\over {k^2}}\right) }^{(1/2) - \omega} \Theta (k^2 - 4 \mu^2 )
\end{eqnarray}
Expression (\ref{21}) equals the phase space of one scalar particle going into two with a $\Phi^3$ dynamics in $d$ dimensions and with a mass $\mu$. We see therefore that at the threshold $k^2 \sim 4 \mu^2$ the cutting rules are satisfied with a $\Phi^3$ dynamics: higher string levels cannot contribute since they are too high in energy, as well as the second cut, and only two scalar particles can be exchanged, with a field theory-like vertex in first approximation.

We will try now to isolate also the discontinuity across the second cut. We will use the same trick as before, focusing this time on the small $t$ region. We analyse the leading behaviour at small $t$
\begin{eqnarray}
\label{22}
\int_0^1 d\nu \, \int_0 dt \, {\left( {{t}\over {2 \pi \alpha '}}\right) }^{2 \alpha ' k^2 {{|g|}\over {[g^2 - {(2 \pi {{\alpha '} B)}^2}]}}} e^{k^2 {{{\pi}^2 {{\alpha '}^2 }}\over {|g| \, t}} - {{{\pi}^2 {{\alpha '}} (2 - d)}\over {6 t}}} 
\end{eqnarray} 
We choose as the upper extreme of integration $t = 2 \pi \alpha '$ and make the change $t \rightarrow {{1}\over {t}}$ to reduce the integral to the already analysed form (\ref{19}). In so doing we also freeze $k$ in the exponent of ${\left( {{t}\over {2 \pi \alpha '}}\right) }^{- 2 \alpha ' k^2 {{|g|}\over {[g^2 - {(2 \pi {{\alpha '} B)}^2}]}}}$ to be at the actual threshold: this combination cannot interfere with the singularity coming from the exponential. After redoing the same analysis as above, we find that the discontinuity is proportional to
\begin{eqnarray}
\label{23}
{\left( 1 - {{4 |g| m^2}\over {k^2}}\right) }^{1 + {|g|}^2 {{2 - d} \over {3 [g^2 - {(2 \pi {{\alpha '} B)}^2}]}}} \, \Theta (k^2 - 4 m^2 |g|)
\end{eqnarray} 
It is interesting in particular to refer this last analysis to the field theory limit. We know that this higher cut embraces $\Re k^2 > 4 m^2 |g|$. When we perform the Seiberg-Witten limit the quantity $4 m^2 |g|$ goes to zero, and the branch cut starts form $k^2=0$ in the field theory, as we can easily see from Eq.(\ref{8}). This singularity is a feature which was generally found in the purely noncommutative field-theoretical approaches even in the magnetic case, where a decoupling (Seiberg-Witten) limit does exists: in \cite{bgnv}, namely in NC gauge theories, it was accompanied in the magnetic case by a cut starting from zero, due to noncommutative massless gluons, while in \cite{gm} it was well separated from the cut due to massive scalars. If we consider this last case, the formal recovering of the cutting rules by implementation of a tree level phase space with noncommutative dynamics did not reveal the nature of these singularities. Our string theory analysis reveals a fact that was already pointed out in the earlier investigations (see for example \cite{mvs}), namely that they have to be related to the closed string sector, which in the usual on-shell analysis manifests itself as poles in the small-$t$ corner. The two point function needs however an off-shell analysis, as we have diffusely pointed out above, in order to relate it to the natural analysis in field theory, but which is aside from the standard domain of perturbative string theory. We identify this fact, namely that the anomalous field theory singularity can be put in correspondence with the problematic operation of going off-shell at the string perturbative level, as the key in the interpretation of the analytic structure of the noncommutative two point function. 

We notice also that for $B=0$ the two branch cuts are superimposed even for $d<2$, and the open string metric coincide with the closed one. In this situation one should instead perform the effective field theory limit keeping this metric fixed, and one would obtain the commutative $\Phi^3$ theory.  
 
\section{Conclusions}
We have shown how the breakdown of perturbative unitarity in noncommutative electric-type field theories can be related, from the point of view of cutting rules, to the appearance of a tachyonic branch cut in the corresponding string theory amplitude when the electric field overcomes the critical value. We have analysed the simple case of a two dimensional brane-worldvolume which is effectively described by a massless scalar noncommutative $\phi^3$ theory. The string amplitude, when viewed as a function of the external squared momentum in the spirit of the off-shell continuation of Di Vecchia et al. (\cite{div1,div2,div3}, see also \cite{bern1,bern2,bern3}) and \cite{bcr}), below the critical field has two branch cuts, both positive, but the zero-slope limit forces the electric field to overcome its critical value. At the same time the quantity that parameterizes one of the branch cuts becomes negative, and the amplitude enters the region of instability. The crucial point is that this forced limit exactly coincides with the amplitude derived from a noncommutative time-like field theory action. The corresponding noncommutative field theory violates unitarity in this situation. 

This is in our opinion a remarkable correspondence. Furthermore, when the magnetic case is reconsidered in the light of our kind of analysis, one realizes new insight into the problem that, in spite of the existence of the Seiberg-Witten decoupling limit, the magnetic theory presents anomalous singularities coming from the closed string sector, as originally pointed out by \cite{mvs}. We have therefore studied the discontinuities of the full string theory amplitude across the branch cuts, keeping the electric field below the critical value in order to have a stable string, and then we have related them to what one finds in the effective field theory. These two branch cuts have different nature, and, in dimensions of the brane-worldvolume different from two, start from different points. This separation, furnished by the $B$-field, allows us to put ourselves in the approximation of an incoming momentum just above the lower threshold, the upper one, as well as the higher string levels, being not still active. In this situation we have recovered, for a small region beyond the lower cut, the cutting rules for simple $\phi^3$ phase space. We have also discussed the features of the upper threshold, which in the case of $B=0$ would result superimposed to the lower one: even in the magnetic case it is related in the Seiberg-Witten limit to the anomalous singularity mentioned above and starting from zero momentum, and in the full string theory amplitude it is driven by the closed string metric. We have discussed the extent to which it is related to an off-shell manifestation of the closed string sector, corresponding to the problematic continuation in $p^2$ already remarked.    

The perspectives opened are, first, to apply our analysis to amplitudes among external states belonging to higher levels in the string spectrum, such as the vectors \cite{bcr,gkmrs}, in order to interpret the result of \cite{bgnv}. In this case, the tachyon contribution of the bosonic open string can be treated with an \it ad hoc \rm subtraction. Next, to get rid of the tachyon directly from the spectrum, one has to pass to the case of superstrings.
\section{Acknowledgements}
I wish to thank Antonio Bassetto for useful discussions and invaluable collaboration during all the work, and for reading the manuscript. I wish to thank Loriano Bonora for reading the first version of the manuscript and for suggestions. I wish to thank C. S. Chu, J. Gomis and S. Kar for very useful suggestions.

\end{document}